\documentclass[aip,rsi, amsmath,amssymb,reprint]{revtex4-1}

\usepackage{graphicx}
\usepackage{dcolumn}
\usepackage{bm}
\usepackage[sort&compress]{natbib}
\usepackage{xspace}
\usepackage{epstopdf}
\usepackage{amsfonts}
\usepackage{epsfig}
\usepackage{textcomp}
\usepackage{changepage}
\usepackage{xcolor}

\newcommand{\mum}{\text{\textmu m}}
\newcommand{\SiN}{\text{Si}_3\text{N}_4}

\begin{document}
\title{Tuning and Stabilization of Optomechanical Crystal Cavities Through NEMS Integration}
\author{Karen E. Grutter}
\affiliation{Laboratory for Physical Sciences, College Park, MD 20740}
\affiliation{Center for Nanoscale Science and Technology, National Institute of Standards and Technology, Gaithersburg, MD 20899-6203}
\author{Marcelo I. Davan\c{c}o}
\affiliation{Center for Nanoscale Science and Technology, National Institute of Standards and Technology, Gaithersburg, MD 20899-6203}
\author{Krishna C. Balram}
\affiliation{Center for Nanoscale Science and Technology, National Institute of Standards and Technology, Gaithersburg, MD 20899-6203}
\affiliation{Maryland Nanocenter, University of Maryland, College Park, MD 20742}
\affiliation{Department of Electrical and Electronic Engineering, University of Bristol, Clifton BS8 1UB, United Kingdom}
\author{Kartik Srinivasan}
\affiliation{Center for Nanoscale Science and Technology, National Institute of Standards and Technology, Gaithersburg, MD 20899-6203}

\begin{abstract}
Nanobeam optomechanical crystals, in which localized GHz frequency mechanical modes are coupled to wavelength-scale optical modes, are being employed in a variety of experiments across different material platforms.  Here, we demonstrate the electrostatic tuning and stabilization of such devices, by integrating a $\SiN$ slot-mode optomechanical crystal cavity with a nanoelectromechanical systems (NEMS) element, which controls the displacement of an additional ``tuning'' beam within the optical near-field of the optomechanical cavity.  Under DC operation, tuning of the optical cavity wavelength across several optical linewidths with little degradation of the optical quality factor ($Q\approx10^5$) is observed.  The AC response of the tuning mechanism is measured, revealing actuator resonance frequencies in the 10~MHz to 20~MHz range, consistent with the predictions from simulations. Feedback control of the optical mode resonance frequency is demonstrated, and alternative actuator geometries are presented.

\end{abstract}

\maketitle
\section{Introduction}

Nanobeam optomechanical crystal cavities have been used in a number of recent experiments in cavity optomechanics.  By co-localizing a high frequency (GHz) mechanical mode with a wavelength-scale optical mode and using phononic and photonic bandgaps to limit mechanical dissipation due to anchors and optical loss due to in-plane radiation, respectively, these systems offer a number of advantages.  Recent experiments utilizing this platform include ground state cooling of the mechanical resonator~\cite{chan_laser_2011}, coupling of the optomechanical cavity to propagating phonons~\cite{balram_coherent_2016,fang_optical_2016}, observation of non-classical correlations between single phonons and single photons\cite{riedinger_non-classical_2016}, and the demonstration of a path towards absolute thermometry based on comparing thermally-driven motion of the mechanical resonator to quantum back-action driven motion~\cite{purdy_quantum_2017}. Moreover, after their initial development and optimization in silicon\cite{chan_laser_2011,chan_optimized_2012}, nanobeam optomechanical crystals have also been demonstrated in $\SiN$~\cite{purdy_quantum_2017,davanco_si3n4_2014,grutter_si3n4_2015}, AlN~\cite{fan_aluminum_2013,bochmann_nanomechanical_2013}, GaAs~\cite{balram_coherent_2016,balram_moving_2014}, diamond~\cite{burek_diamond_2016}, and LiNbO$_3$~\cite{liang_high-quality_2017}.  

\begin{figure*}[!htbp]
\centering
\includegraphics[width=\linewidth]{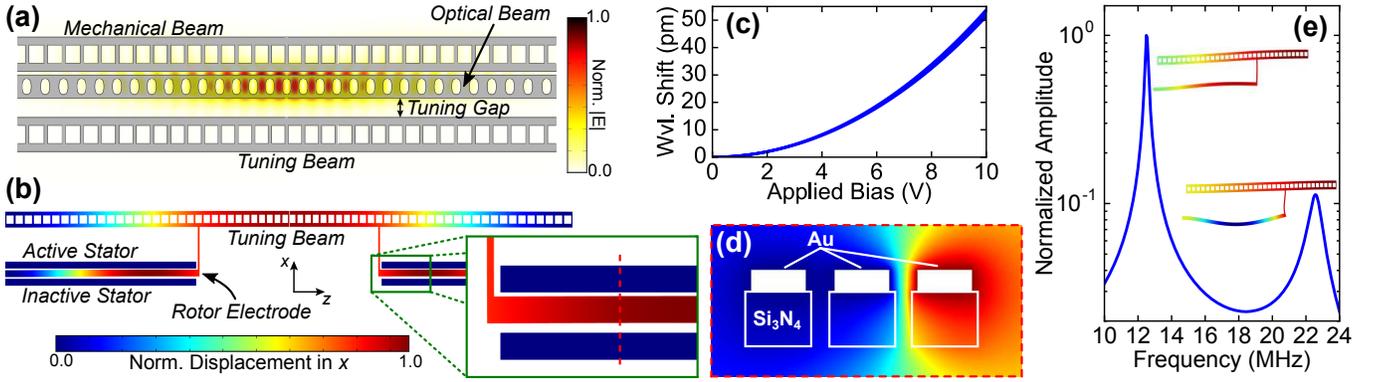}
\caption{(a) Finite element method (FEM) simulation of the fundamental optical slot mode in the presence of the third ``tuning'' beam placed far enough away from the optical beam that it does not support a slot mode.  (b) FEM simulation of the displacement of the tuning beam attached to NEMS electrode structures. When a bias is applied to the anchored active ``stator'' electrodes while the moving ``rotor'' electrodes are grounded, the rotor electrodes and the attached tuning beam are pulled in the $+x$ direction (the opposite,  inactive stators are also grounded). (inset) Zoomed-in view of the NEMS electrodes in the region indicated by the green box.  (c) The simulated resonant wavelength shift for varying bias applied to the NEMS actuators, assuming gold electrodes 100~nm thick on top of $\SiN$ and an initial tuning gap of 200~nm. This graph was derived by combining the optical and electro-mechanical FEM simulations; details can be found in Appendix~\ref{app:sim}. (d) The electric potential of the cross section of the electrodes shown in (b), where the bias is applied to gold layer on top of the $\SiN$ structure.  (e) Simulated frequency response of NEMS-driven tuning beam, assuming an isotropic loss factor in the $\SiN$ of 0.017. Insets show the two symmetric eigenmodes in the frequency range 10~MHz to 24~MHz.}
\label{fig:design}
\end{figure*}

One avenue for further development of these devices is through their integration with other elements that provide additional mechanisms to help control and utilize optomechanical interactions.  For example, there have been recent demonstrations connecting nanobeam optomechanical crystals to GHz radio frequency waves, using piezoelectric media~\cite{bochmann_nanomechanical_2013,balram_coherent_2016}.  Here, we consider the integration of Si$_3$N$_4$ nanobeam optomechanical crystals with nanoelectromechanical systems (NEMS)~\cite{midolo_nano-opto-electro-mechanical_2018}, and in particular, a NEMS actuator that enables electrostatic tuning and feedback stabilization of the optical cavity resonance.  Our NEMS actuator provides a combination of DC tuning of the optical cavity mode wavelength over a range of a few cavity linewidths, and AC modulation of the wavelength with a bandwidth of $\approx$10~MHz.  We demonstrate feedback control of the optical cavity resonance, and explore the tradeoff between larger tuning range and lower modulation bandwidth.  In comparison to previous work on electrostatically-tunable optomechanical systems, including photonic crystals~\cite{frank_programmable_2010,winger_chip-scale_2011,chew_nanomechanically_2011,miao_microelectromechanically_2012,pitanti_strong_2015,du_mechanically-tunable_2016}, here the actuator is entirely independent of the optomechanical resonator, so that the advantageous properties of the nanobeam optomechanical crystal system are retained.  This incorporation of high-bandwidth electrical tuning and control of the optical resonance can enable stabilization of the intracavity optical power for long-term measurements, e.g., in the presence of thermal fluctuations.

\section{Design}
\label{sec:design}

To enable NEMS integration without affecting the properties of the optomechanical cavity, we utilize the slot-mode optomechanical crystal geometry demonstrated in Ref.~\citenum{grutter_slot-mode_2015}. This geometry (Fig.~\ref{fig:design}a) confines the optical mode in the narrow slot region between two patterned nanobeams, one of which (the `optical beam') provides lateral confinement (i.e., along the $z$-direction) of the optical mode, and the other of which confines a GHz frequency mechanical breathing mode (the `mechanical' beam).  This tight modal confinement within the slot region enhances optomechanical interactions, because the optical resonance frequency sensitively depends on the size of the slot between the two beams, which is modulated by the breathing mode vibrations.  It is also particularly well-suited for the tuning strategy we employ, because the strong modal confinement allows placement of metal electrodes relatively nearby without causing optical loss. In addition, the mechanical breathing mode can interact with additional elements on its free side, enabling multi-mode applications such as optical wavelength conversion~\cite{grutter_slot-mode_2015}.  We note that this strategy of NEMS actuation could also be compatible with a single-nanobeam optomechanical crystal, but it would require additional design considerations to minimize degradation of the optical performance.

In this structure, the optical resonance wavelength $\lambda_0$ could be shifted by directly changing the slot width, but this would also intrinsically change other important device characteristics, including the optomechanical coupling rate and the optical quality factor.  Thus, in order to independently control $\lambda_0$, we add another ``tuning'' beam to the opposite side of the optical beam, as shown in Fig.~\ref{fig:design}(a).  The tuning gap is large enough that it does not support another slot mode, but it does perturbatively interact with the evanescent tail of the optical slot-mode, tuning the mode effective index ($n_{\textup{eff}}$) and thereby shifting the optical resonance.  In order to minimize scattering losses, the tuning beam geometry is identical to the mechanical beam.

We design NEMS capacitive transducers to attach to the tuning beam, as shown in Fig.~\ref{fig:design}(b), which in turn control the width of the tuning gap. In this design, the moving ``rotor'' electrodes are attached to the outer side of the tuning beam via long (1.895~\mum) tethers in order to keep the metal electrodes away from the optical mode.  In addition, the connection point of the rotor along the length of the tuning beam is optimized to minimize optical scattering.  The direction of actuation of the tuning beam can be either toward or away from the optical slot mode, depending on which anchored ``stator'' electrodes are active.  In the simulation of Fig.~\ref{fig:design}(b), the active stators pull the tuning beam in the $+x$ direction, while the opposite stator electrodes are grounded, like the rotors, and thus do not exert force in the $-x$ direction.  Combining finite element method (FEM) simulations of $\lambda_0$ at different tuning gaps with simulations of the displacement of the tuning beam for different biases applied to the NEMS actuator, we can predict the tuning range of this design.  For an initial tuning gap of 200~nm and initial rotor-stator electrode gaps of 130~nm, these simulations suggest $\lambda_0$ could be shifted by about 50~pm with an applied bias of 10~V (Fig.~\ref{fig:design}(c)).

Active tuning of $\lambda_0$ enables on-chip locking of the optical cavity to a pump laser, which may be especially important for counteracting thermal effects. To determine the possible tuning speed, we simulated the frequency response of the NEMS-actuated tuning beam, which shows symmetric mechanical modes at about 13~MHz and 23~MHz (Fig.~\ref{fig:design}(e)). This satisfies our goal to design the NEMS actuator with fundamental frequency higher than the effective frequency of the cavity thermal response, which is expected to be in the MHz range for suspended $\SiN$ structures, an order of magnitude lower than the simulated NEMS frequency.

\begin{figure*}[!htbp]
\centering
\includegraphics[width=0.9\linewidth]{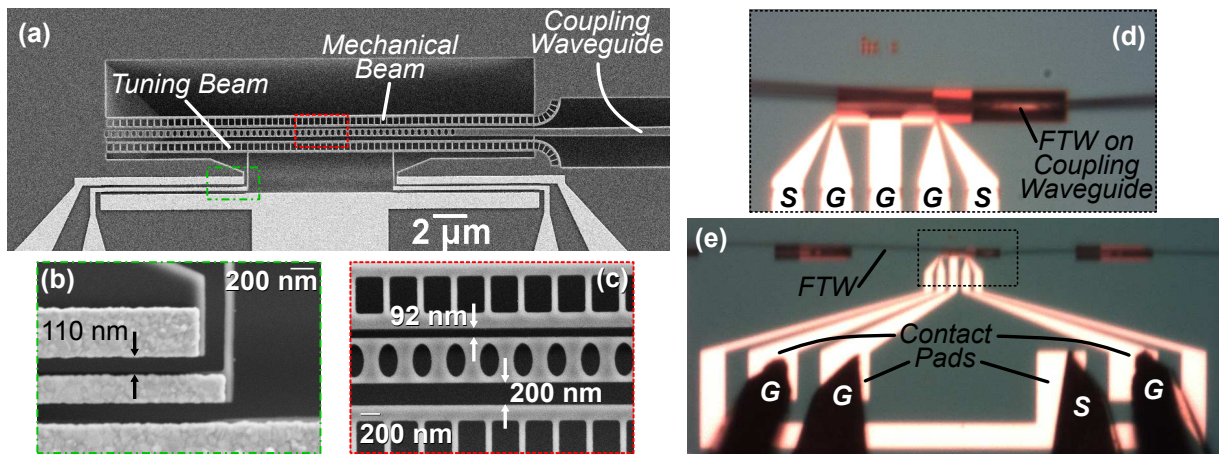}
\caption{(a) Scanning electron microscope (SEM) image of fabricated device.  (b) Zoomed-in SEM image of one of the capacitive transducer structures (green dashed box region in (a)).  (c) Zoomed-in SEM image of the center of the slot mode structure (red dashed box region in (a)).  The slot between the mechanical beam and the optical beam is about 92~nm wide, and the gap between the optical beam and the tuning beam (the `tuning gap' in Fig.~\ref{fig:design}(a)) is about 200~nm wide. (d) Device under test, with optical fiber taper waveguide (FTW) resting on coupling waveguide. (e) Optical microscope image of fabricated device on characterization setup, including FTW and electrical probes ($G$~=~ground, $S$~=~signal).  Dashed box indicates region shown in (d).}
\label{fig:device}
\end{figure*}

Adding a third nanobeam to the slot-mode device complicates the method for coupling light into the optical cavity, as the tuning beam is located at the ideal in-plane location for evanescent coupling to the slot mode.  In previous demonstrations of triple nanobeam devices, evanescent coupling was accomplished by instead hovering a fiber taper waveguide (FTW) directly over the slot\cite{grutter_slot-mode_2015}, but this adds another source of instability to the system. Thus, for this device, we modified the design slightly to enable end-fire coupling via an on-chip waveguide into the slot mode. This on-chip coupling waveguide has at its input a tapered coupler designed for directional coupling to a FTW resting on it. Details can be found in Appendix~\ref{app:endfire}.

\section{Fabrication}

Devices were fabricated in 250~nm thick stoichiometric $\SiN$ deposited via low-pressure chemical vapor deposition on Si.  Before defining the nanobeams, $\approx 65$~nm thick Au electrodes, with an $\approx 10$~nm thick Cr adhesion layer, were defined using electron-beam lithography and lift-off.  Then, the $\SiN$ features were patterned (aligned to the electrodes), etched, and released as previously described, using electron-beam lithography, reactive ion etching, and KOH\cite{grutter_slot-mode_2015}.  The fabricated NEMS actuators include stator electrodes on both sides of the rotor electrode, designed for actuation of the tuning beam both toward and away from the optical cavity.  For single-direction actuation, both the rotor and the unused stator electrodes are grounded.

Fabricated slots between optical and mechanical beams range from about 70~nm to about 90~nm. Tuning gaps for all devices are about 200~nm.  The released distance between the NEMS actuator's electrodes ranges from about 90~nm to 140~nm.  These measured post-release dimensions differ somewhat from the lithographically-defined pattern due to the tensile stress intrinsic to the stoichiometric $\SiN$ film.  Scanning electron microscope (SEM) images of a fabricated device are shown in Figure~\ref{fig:device}.

\section{Large-Signal Characterization}

Devices were characterized optically at room temperature and atmospheric pressure with a 980~nm external cavity tunable diode laser, which was coupled into the devices via a dimpled FTW touching the coupling waveguide, as described in Appendix~\ref{app:endfire}. The FTW was held in place by Van der Waals forces.  The NEMS actuators were driven via an RF probe on the contact pads.  The driving signal was applied to the stator electrodes pulling the tuning beam toward the optical cavity, while the rotor and inactive stator electrodes were grounded, as shown in Figure~\ref{fig:device}(d) and (e). The vibrations of the mechanical beam and tuning beam were detected in the modulation of the optical signal output.

\begin{figure*}[htbp]
\centering
\includegraphics[width=\linewidth]{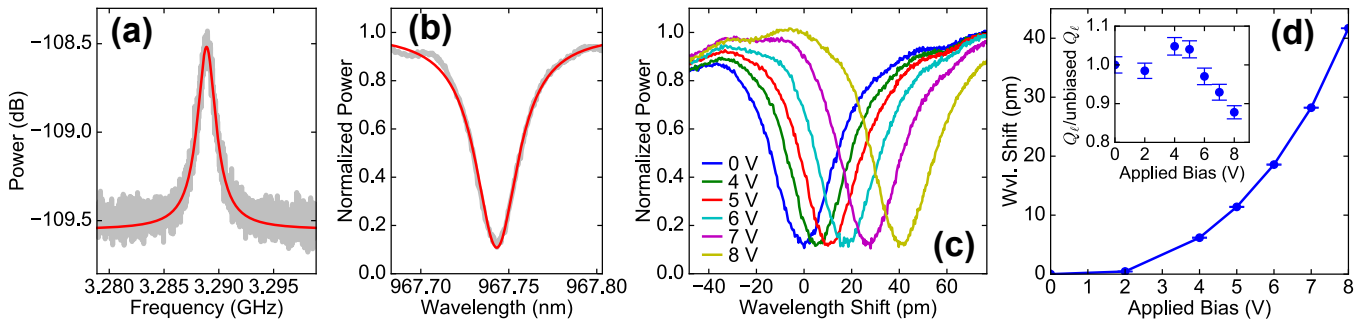}
\caption{(a) Mechanical breathing mode centered at $3.289~\text{GHz}~\pm~0.007~\text{GHz}$ with quality factor $Q_m=1680~\pm~40$.\cite{mecherrornote} This power spectral density plot is normalized to a power of 1~mW~=~0~dB. The Lorentzian fit of the spectrum is in red. (b) Fit of normalized optical slot mode with no applied bias, $Q_i = 100000~\pm~6000$.\cite{mecherrornote}  (c) Series of optical spectra for different applied biases, with the NEMS actuator pulling the tuning beam toward the optical slot mode.  (d) Shift of optical resonance as a function of applied bias. (inset) Loaded quality factor $Q_{\ell}$ as bias changes. Error bars represent the 95~\% confidence interval of the Lorentzian fit of the spectrum at each bias value.}
\label{fig:largesignal}
\end{figure*}

The measurements presented in this work are from the device design shown in Figure~\ref{fig:device}.  The spectrum of the mechanical breathing mode at $\approx 3.3$~GHz with 0~V applied to the NEMS actuator is shown in Figure~\ref{fig:largesignal}(a).  The breathing mode quality factor $Q_m=1680~\pm~40$.\cite{mecherrornote} This is slightly lower than the measured $Q_m$ of previous slot-mode optomechanical crystals without tuning beams\cite{grutter_slot-mode_2015}, likely due to the modification of the anchoring conditions of one end of the mechanical beam to accommodate the coupling waveguide.  The optical slot mode at $\approx 968$~nm (Figure~\ref{fig:largesignal}(b)) had an intrinsic optical quality factor of $Q_o=100000~\pm~6000$~\cite{mecherrornote} with 0~V applied to the MEMS actuator, which is comparable to $Q_o$ previously measured in $\SiN$ slot-mode optomechanical crystals\cite{grutter_slot-mode_2015}.  This shows that the presence of the tuning structure does not significantly adversely perturb the optical mode.

We characterized the performance of the NEMS actuator by applying a varying DC bias to the active stator electrodes, pulling the tuning beam toward the optical cavity, while grounding the rotor and the opposite, inactive stator.  For each applied bias, we measured the optical slot-mode spectrum with the tunable laser, as shown in Figure~\ref{fig:largesignal}(c). To ensure accuracy of the wavelength tuning measurement, we sent 10~\% of the input laser power to a wavemeter for a concurrent wavelength measurement.  By applying an 8~V DC bias to the NEMS actuator, we shifted $\lambda_0$ by $41.69~\text{pm} \pm 0.05~\text{pm}$, more than quadruple the intrinsic linewidth of the optical mode. This shift is consistent with the simulated wavelength shift, shown in Fig.~\ref{fig:design}(c).  The loaded optical quality factor $Q_\ell$ changes very little (within $\approx\pm10~\%$ of $Q_\ell$ at $V=0$~V) over this tuning range, which shows that the NEMS actuator controls $\lambda_0$ without introducing a significant parasitic optical loss channel.

\section{Small-Signal and Locking Performance}
To determine the small-signal performance of the NEMS actuator, we combine the DC voltage drive with an oscillating voltage signal generated by a vector network analyzer (VNA), and read the resultant oscillation on the photodetected optical signal output when the laser wavelength is fixed on the blue-detuned shoulder of the optical resonance. The magnitude of the NEMS-induced oscillation increases as the DC bias increases, both because the tuning beam is closer to the optical slot mode and because the electrodes of the NEMS actuator are closer together.  This S21 measurement essentially characterizes the slope of the $\lambda_0$-vs-bias graph (Fig.~\ref{fig:design}(c)) at a fixed DC bias. For a DC bias of 0~V, the slope is zero, and thus the S21 spectrum should show no response from the actuation of the tuning beam.  Increasing the DC bias increases the slope, and thus the magnitude of the optical transmission oscillation due to the tuning beam increases.

\begin{figure*}[htbp]
\centering
\includegraphics[width=\linewidth]{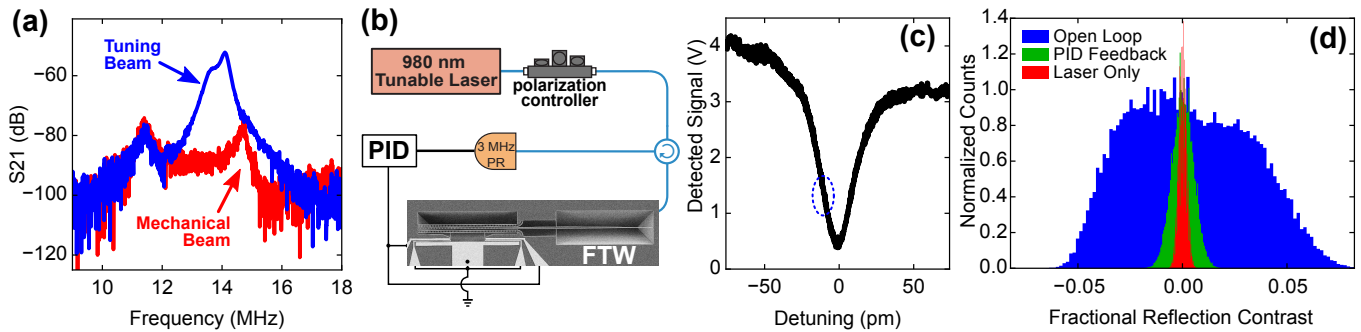}
\caption{(a) S21 measurement (RF electrical drive input and photodetected output) of the NEMS-actuated slot mode optomechanical crystal at 0~V DC bias (red) and 10~V DC bias (blue), where the probe laser is fixed onto the blue-detuned shoulder of the optical cavity mode.  (b) Characterization setup for proportional-integral-derivative (PID) controller locking of the optomechanical cavity to laser.  (c) The optical spectrum of the resonance used for the locking experiment. Circled in blue is the point at which the laser was set for locking.  (d) A comparison of the reflected signal variation over one minute with laser wavelength fixed at the detuning indicated in (c), for open loop (blue), PID feedback into NEMS actuator (green), and laser going directly into the detector (red).}
\label{fig:smsig}
\end{figure*}

This S21 measurement at two different DC bias values is shown in Figure~\ref{fig:smsig}(a). The detected oscillation of the optical signal is a combination of the tuning beam response and the coherently-driven motion of the mechanical beam of the optomechanical crystal.  In addition to the $\approx 3.3$~GHz mechanical breathing mode, the mechanical beam also supports clamp-clamp Euler beam modes which couple strongly enough to the optical slot mode that the optical power threshold for self-oscillation is very low~\cite{grutter_slot-mode_2015}. The lowest order of these beam modes is expected to be close in frequency to the tuning beam fundamental frequency, because the beams' geometry is the same except for the attached NEMS actuator and any fabrication variation.  In the S21 measurement, for 0~V bias applied to the NEMS actuator, we see only the spectrum due to self-oscillation of the mechanical beam, with the lowest-order beam resonance around 14.7~MHz.  As the NEMS bias is increased, an additional peak emerges around 13.7~MHz due to the response of the tuning beam. This peak corresponds well to the simulated frequency response shown in Fig.~\ref{fig:design}(e).

In order to lock the cavity resonance $\lambda_0$ to the fixed laser wavelength, we use the detected optical signal as the input to a proportional-integral-derivative (PID) controller (Fig.~\ref{fig:smsig}(b)). The output signal of the PID controller is applied to the NEMS actuator. With the laser wavelength initially set to the blue shoulder of the optical resonance (indicated in Fig.~\ref{fig:smsig}(c)), we adjusted the PID controller to keep the magnitude of the detected optical signal constant.  The optimal parameters we found were as follows: $P=0.738$, $I=3.036\times10^5$~s$^{-1}$, and $D=1.179~\textup{\textmu s}$. With the PID feedback enabled, we measured an optical signal stability improvement of a factor of $7.21~\pm~0.02$ in comparison to the full-width half-maximum of the free-running optical signal over a 1~min integration time (Fig.~\ref{fig:smsig}(d)). We also measured the laser intensity variation alone, which was $38~\%\pm2~\%$ of the PID-controlled fluctuation and $5.3~\%\pm0.3~\%$ of the open-loop fluctuation. (Error in fluctuation comes from the Fig.~\ref{fig:smsig}d histogram bin width.)

Because the laser fluctuation is much less than the open-loop device transmission variation, we assume that the majority of the open-loop fluctuation is due to instability of the device itself.  This shows that the NEMS actuator is an effective tool for locking $\lambda_0$ in an unstable optomechanical cavity to a fixed laser, and is capable of nearly achieving laser-limited stability.

\section{Discussion}
\label{sec:discuss}

The addition of a third nanobeam placed in the evanescent field of a slot-mode optomechanical crystal structure provides a convenient interface to a NEMS actuator for electrical control of the optical slot-mode. Using this structure, we have demonstrated a resonance shift of $41.69~\text{pm}~\pm~0.05~\text{pm}$ for a DC bias of 8~V, corresponding to a shift greater than four cavity linewidths. Unlike previous examples of capacitive-transducer-tuned slot-mode photonic crystals\cite{chew_dynamic_2010,pitanti_strong_2015}, this strategy shifts the optical resonance without directly tuning the other optomechanical characteristics of the cavity under test, and the cavity-tuning transducer is separate from the mechanical resonator.

External control of the optical resonance enables active locking of the cavity resonance to a fixed laser.  This is especially important for stabilizing optomechanical cavities in long-term measurements, a significant challenge that has been previously addressed by instead actively tuning the laser to follow the fluctuations of the cavity. We have shown that, using a simple PID controller to drive the NEMS actuator, we can improve the cavity stability by a factor of $7.21~\pm~0.02$ in comparison to the unlocked optical cavity.  For this demonstration, we used the cavity transmission as the input to the PID controller, but alternative feedback signals, such as the oscillation amplitude in the mechanical mode of interest\cite{huang_direct_2017}, are also compatible with this technique.

\begin{figure*}[htbp]
\centering
\includegraphics[width=\linewidth]{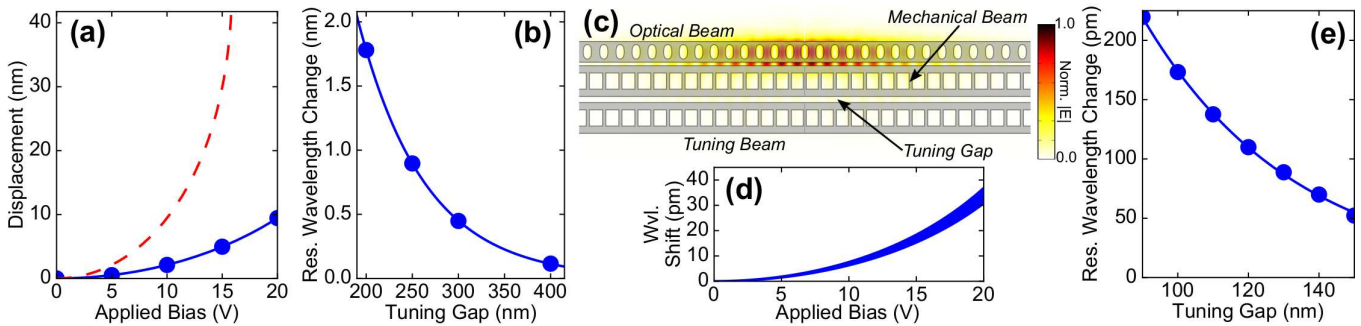}

\caption{(a) Simulated beam center displacement with respect to applied bias for the structure shown in Fig.~\ref{fig:design}(b). The line indicates the fit of the results to analytical parallel-plate actuator equations. Red dashed line corresponds to a softer spring constant $k/4$. (b) Simulated optical resonance wavelength shift with respect to gap between tuning beam and optical beam for the device shown in Fig.~\ref{fig:design}(a). (c) FEM simulation of the fundamental optical slot mode in the presence of the third ``tuning'' beam placed adjacent to the mechanical beam (``TMO''). (d) The simulated resonant wavelength shift for varying bias applied to the NEMS actuator, using the same electro-mechanical FEM simulation as in (a), optical FEM simulations from (c), and assuming an initial tuning gap of 110~nm. (e) Simulated optical resonance wavelength shift with respect to gap between tuning beam and mechanical beam for TMO device.}

\label{fig:discuss}
\end{figure*}

The optical resonance tuning range demonstrated in this work is not particularly wide in comparison to other demonstrations of NEMS- and MEMS-actuated optical tuning~\cite{frank_programmable_2010,winger_chip-scale_2011,chew_dynamic_2010,miao_microelectromechanically_2012,pitanti_strong_2015}.  However, the goal with this work is to enable locking fast enough to compensate for thermal fluctuations of the device, which necessitates optimization away from large NEMS displacement. (For an ideal parallel-plate capacitive transducer, the mechanical frequency $\Omega_m\propto\sqrt[]{k}$ and the required bias to achieve a given displacement $V\propto\sqrt[]{k}$.  Thus, increasing $\Omega_m$ also increases the required bias by the same factor.)  This NEMS design could be modified for a wide optical tuning range simply by reducing the effective spring constant.  In Fig.~\ref{fig:discuss}(a), we show the FEM-simulated displacement of the NEMS actuator with respect to applied bias and fit it to theoretical expressions for a parallel-plate capacitive transducer, detailed in Appendix~\ref{app:sim}.  We also show that same fit function with the spring constant reduced to $1/4$ the original designed value.  With this reduced spring constant, the actuation would increase to $\approx 40$~nm displacement with an applied bias of 18~V, corresponding to a resonance shift of over 500~pm (see Fig.~\ref{fig:discuss}(b)).  This range could be further extended by changing the design of the capacitive transducer itself.  The upper bound on the actuation distance of a parallel-plate capacitive transducer, which approximates the behavior of this design, is the pull-in distance, which is $\approx 1/3$ the initial electrode gap.  However, a much longer actuation distance is possible using alternative designs, such as the electrostatic comb drive\cite{tang_laterally_1989}. Both softening the spring and moving to a larger capacitive transducer structure would inherently decrease the mechanical frequency of the actuator, but this tradeoff may be favorable for certain applications.

One of the key features of this NEMS-actuator based control is that integration with the optomechanical crystal in the fabrication process is straightforward. No modifications to the processing of the optomechanical device layer itself were required; patterning and deposition of the NEMS electrodes were simply an added layer.  As a result, similar NEMS structures could be easily integrated with other device designs. 

As a simple example, instead of placing the tuning beam adjacent to the optical beam of a slot-mode optomechanical crystal, we can place it adjacent to the mechanical beam while attached to the same NEMS actuator design (Fig.~\ref{fig:discuss}c). This ``TMO'' structure leaves one side of the optical beam available. Because the tuning beam is farther away from the optical slot mode, its effect on the optical resonance is not as strong as in the ``TOM'' device characterized in this work (at a 10~V bias, the simulated optical resonance shift is about 20~\% of that of the TOM device), but the exposed optical beam simplifies coupling to the optical mode.  As a result, no modification of the beam anchoring to enable end-fire coupling is required.  Preliminary measurements of a prototype of this device, shown in Appendix~\ref{app:tmo}, indicate that high mechanical and optical quality factors can be achieved with the TMO geometry.

Finally, the addition of this electrical input/output channel could open the door for new measurements involving the interaction of optical, mechanical, and electrical/RF domains. Previous work has demonstrated this multi-domain interaction in piezoelectric materials\cite{balram_coherent_2016,bochmann_nanomechanical_2013} and doped semiconductors\cite{beyazoglu_multi-material_2014,sridaran_electrostatic_2011}, but the additive fabrication process of this work separates the electrical component design and fabrication from the optomechanical components, thereby expanding the set of possible material systems for these studies. One specific application is in microwave-to-optical transduction mediated by mechanics, where the motion of a mechanical element is coupled to both an optical resonator and a microwave resonator.  In particular, the device concepts presented here can be extended to address the scheme proposed in Ref.~\citenum{Davanco_2012}, in which a slot mode optomechanical crystal~\cite{grutter_slot-mode_2015} is integrated with a planar cavity electromechanical system~\cite{Fink_2016}.  The ability to combine slot mode optomechanics with electrodes that are separated by a narrow tuning gap, as demonstrated in this work, is critical to the realization of such a platform.

\section*{Acknowledgements}
K. Grutter acknowledges the NIST/NRC postdoctoral fellowship program. This work was partly supported by the DARPA MESO program.

\appendix
\section{Fitting Optical Tuning Simulations}
\label{app:sim}
The graphs of the simulated resonance wavelength shift with respect to applied bias (Figures~\ref{fig:design}(c) and \ref{fig:discuss}(d)) are a combination of the NEMS actuator simulation results and the optical simulation results. To combine these results, we fit the separate simulations to theoretical curves and use those fitting functions together to form an expression for $\Delta \lambda(V)$.

The theoretical behavior of a parallel-plate capacitive transducer is well-known~\cite{senturia_microsystem_2007}. The forces on the moving rotor electrode are the force due to the electrodes $F_e$ and the spring force due to its suspending structure $F_s$:
\begin{subequations}
  \begin{equation}
	F_e\left(\Delta x, V\right) = \frac{\epsilon_0 A}{2 \left(g_{e0}-\Delta x\right)^2}V^2
  \end{equation}
  \begin{equation}
  	F_s\left(\Delta x\right) = -k~\Delta x
  \end{equation}
\label{eq:nems}
\end{subequations}

\noindent Here, the effective electrode gap at zero applied bias $V$ is $g_{e0}$, the displacement is $\Delta x$, the effective cross-sectional area of the capacitive transducer is $A$, and the effective spring constant of the suspending structure is $k$.  The net force on the rotor electrode is $F_{\text{net}}\left(\Delta x, V\right) = F_e\left(\Delta x, V\right) + F_s\left(\Delta x\right)$.  To determine $\Delta x(V)$, $F_{\text{net}}$ is set to 0, and we analytically solve the resulting cubic equation for $\Delta x$. For $\Delta x > g_{e0}/3$, there are no equilibrium solutions to this equation because the capacitive force increases faster than the spring force with respect to displacement.  This is the ``pull-in'' condition.

\begin{figure*}[htbp]
\centering
\includegraphics[width=\linewidth]{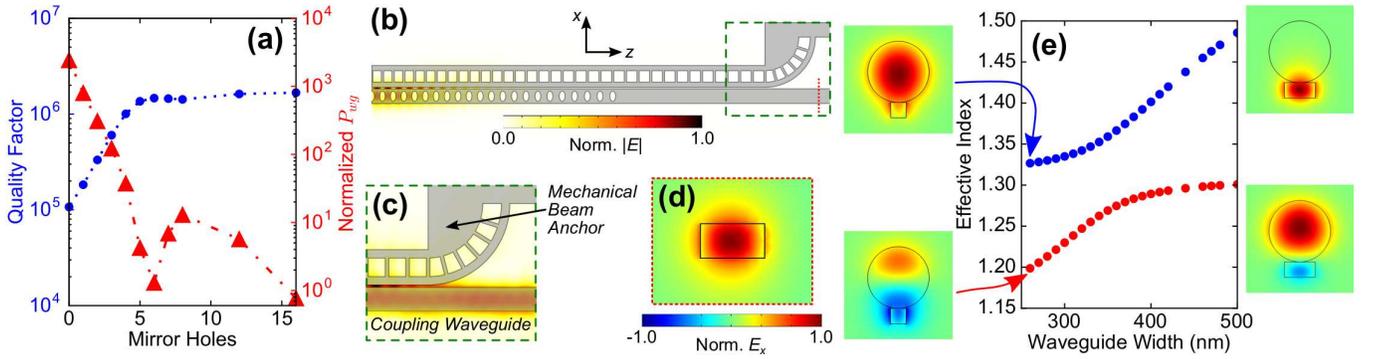}
\caption{(a) Results of FEM simulation of the optical slot mode for varying number of holes in one half of the optical beam. Optical quality factor is shown in blue circles, and the power flow through the coupling waveguide boundary is shown in red triangles. The waveguide power is normalized by the power flow through the other end of the optical beam, which has all 16 of the original mirror holes. (b) FEM of optical slot mode with number of mirror holes on coupling side reduced to 3 and curved, spoked mechanical beam anchor designed for minimal scattering in the transition between the slot mode and the coupling waveguide. (c) Zoomed-in image of the electric field magnitude around the curved mechanical beam anchor, rescaled to the maximum $|E|$ value in the zoomed-in area. (d) Cross section of the coupling waveguide from the simulation in b, showing $E_x$. (e) FEM simulations of the effective index of the symmetric (blue) and anti-symmetric (red) supermodes of a $1~\mum$ diameter FTW touched down on a $\SiN$ waveguide at a range of waveguide widths.  To the sides of the graph are the corresponding simulated $E_x$ field profiles.}
\label{fig:coupling}
\end{figure*}

Although the NEMS structure of this work is not an ideal parallel-plate capacitive transducer, its behavior with respect to applied bias can be approximated by these expressions.  We fit the FEM simulation of the tuning beam displacement with respect to applied bias for the NEMS structure to Eq.~\ref{eq:nems}, with the fitting parameters $g_{e0} = 140.07$~nm~$\pm$~0.02~nm and the ratio $A/k = 91020~\mum^3/\text{N} \pm 20~\mum^3/\text{N}$ (the uncertainty in these fitting parameters represents the 95~\% confidence interval on the fit.)  The simulation results and this fit are shown in Fig.~\ref{fig:discuss}(a).

To fit the optical simulation results, we assume that, to first order, the relationship between the optical resonance wavelength shift $\Delta\lambda$ and the tuning gap $g_t$ is exponential:

\begin{equation}
\Delta\lambda\left(g_t\right)=\Delta\lambda_0 e^{-\gamma g_{t}}
\label{eq:opt}
\end{equation}

\noindent We thus fit the FEM simulation of the optical resonance shift at different tuning beam positions to Eq.~\ref{eq:opt}, with $\Delta\lambda_0$ and $\gamma$ as fit parameters. The simulation data and fits for two different geometries are shown in Fig.~\ref{fig:discuss}(b) and (e).  For the TOM geometry (Fig.~\ref{fig:design}(a)), $\Delta\lambda_0 = 27.8$~nm~$\pm$~0.4~nm and $\gamma = 0.01375~\text{nm}^{-1}~\pm~0.00007~\text{nm}^{-1}$.  For the TMO geometry (Fig.~\ref{fig:discuss}(c)), $\Delta\lambda_0 = 1.8$~nm~$\pm$~0.1~nm and $\gamma = 0.0231~\text{nm}^{-1}~\pm 0.0005~\text{nm}^{-1}$. (The uncertainty in these fitting parameters represents the 95~\% confidence interval on the fit.)

We re-write the optical fit expression in terms of displacement $\Delta x$ by choosing an initial tuning gap $g_{t0}$ and setting $g_t = g_{t0} - \Delta x$. (This assumes the actuator is arranged such that applying a bias reduces the tuning gap.) We use this combined expression to construct Figures~\ref{fig:design}(c) and \ref{fig:discuss}(d), where the width of the curve incorporates the uncertainty in all of the fit parameters.

\section{End-Fire Coupling Design}
\label{app:endfire}
Adding the tuning beam and NEMS actuator adjacent to the optical beam of the slot-mode optomechanical crystal (Fig.~\ref{fig:design}(a)) blocks evanescent coupling to the side of the optical cavity, thereby eliminating the most straightforward means of optical characterization.  Evanescent coupling from above the cavity can be accomplished by hovering a fiber taper waveguide (FTW) directly over the slot, but the un-anchored FTW adds another source of instability to the system. To avoid this instability, we modified the device design for end-fire coupling via an integrated on-chip waveguide into the slot mode. For ease of characterization, we also optimized this coupling waveguide for directional coupling to a FTW, which is used to deliver signals to/from the chip.

\begin{figure*}[htbp]
\centering
\includegraphics[width=\linewidth]{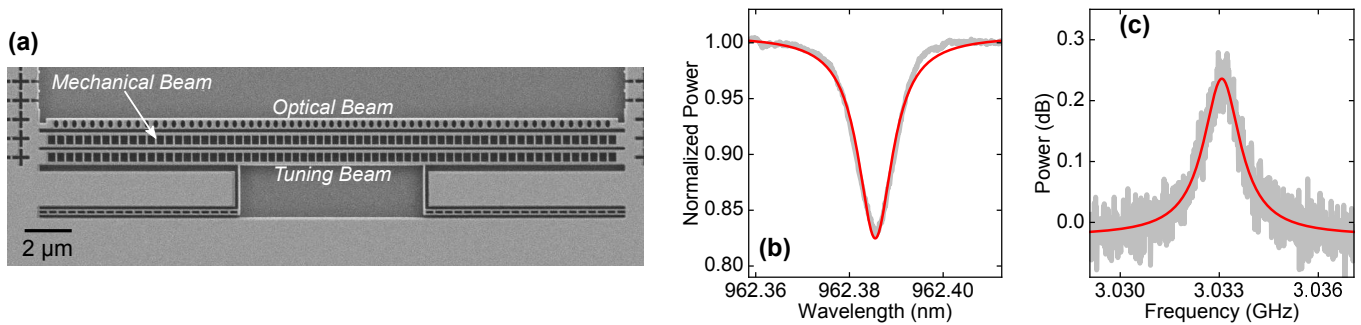}
\caption{(a) SEM image of a fabricated prototype structure in which the ``tuning'' beam is placed adjacent to the mechanical beam (``TMO'') instead of the optical beam. This image was taken prior to the KOH release of the structure, and electrodes were not included in this fabrication process.  (b) Measured optical resonance of a fabricated prototype TMO device, with intrinsic optical $Q_i=120000~\pm~2000$.\cite{mecherrornote} (c) Measured mechanical breathing mode of a prototype TMO structure's mechanical beam, centered at $\approx 3.033$~GHz with quality factor $Q_m=2260~\pm~80$.\cite{mecherrornote}  This power spectral density plot is referenced to a power of 1~mW~=~0~dB.}
\label{fig:tmo}
\end{figure*}

End-fire coupling into the slot mode requires modification of both the optical design and the mechanical anchoring of the beams. Optically, we had to adjust the photonic crystal cavity design to allow light to couple to the cavity from the end of the optical beam.  The photonic crystal nanobeam is designed with a cavity region in the center, where the hole spacing changes, and mirror regions at the ends, where the hole spacing is constant.  By reducing the number of mirror holes on one end of the optical beam, we increase the coupling of light to the cavity from that end. Fig.~\ref{fig:coupling}(a) shows how the simulated optical quality factor of the slot mode decreases as mirror holes are removed. In general, we also see that reducing the number of mirror holes increases the optical power flowing out of the end of the optical beam. (There is a deviation from this trend around five mirror holes, likely due to a localized resonance formed by that particular hole arrangement with respect to the unaltered mechanical beam.)  For our design, we chose three mirror holes to approximately achieve critical coupling to the cavity.

The transition from the slot mode device to the coupling waveguide is also critical; without modification, this introduces a significant optical scattering site. In addition, the transition must include physical anchoring of the mechanical beam.  We designed the mechanical beam to curve away from the coupling waveguide, following a $1.6~\mum$ radius circular path, until it could be anchored away from the coupling waveguide (Fig~\ref{fig:coupling}(b) and (c)).  FEM simulations suggest this design does not significantly affect $Q_o$, and the optical power is confined to the coupling waveguide beyond the transition (Fig.~\ref{fig:coupling}(d)). Further optimization of this transition to minimize optical scattering is possible\cite{palmer_low-loss_2013}, but the most optically-efficient geometries are more challenging to implement while also sufficiently anchoring the mechanical beam.

With these modifications of the optical cavity design for end-fire coupling and the transition to the coupling waveguide mode, this triple nanobeam structure could be easily integrated with other on-chip photonic structures, including grating couplers and edge couplers. In this work, for simple fabrication and characterization, we designed a directional coupler between the on-chip coupling waveguide and an optical fiber taper waveguide (FTW) with a diameter of $\approx 1~\mum$, which is in vertical contact with the on-chip waveguide.  To ensure a smooth transition from the quasi-TE mode of the rectangular waveguide to that of the circular FTW, we simulated the effective index of the symmetric and anti-symmetric optical eigenmodes of the FTW-waveguide structure for different waveguide widths (Fig.~\ref{fig:coupling}e). The anti-crossing point of these modes, around a waveguide width of 360~nm, is indicative that coupling between them can occur if the waveguide width is tapered through this point in the contact area between the FTW and the waveguide.

A finite-difference time domain (FDTD) simulation of a $15~\mum$ long rectangular $\SiN$ waveguide continuously touching a $1~\mum$ diameter circular glass waveguide, with the $\SiN$ waveguide width tapered from 471~nm to 250~nm, shows a $> 90~\%$ transmission from the circular waveguide mode to the rectangular waveguide mode.  Unfortunately, the geometry of the FTW-waveguide contact zone is not as well-controlled in experiment, so the direct applicability of these FDTD simulation results is limited.  However, in fabricated devices, we do see effective coupling.  In a diagnostic slot-mode device, we compare the thermo-optic shift of the optical resonance coupled through the waveguide to that measured while hovering the FTW over the cavity.  Using this method, we infer $\approx 2.5\times$ more power coupled into the optical cavity through the coupling waveguide, showing that this structure enables efficient optical coupling to the slot-mode optomechanical crystal cavity.

\section{Alternative Tunable Device Design}
\label{app:tmo}
The device presented in the main body of this work places a tuning beam in the slot mode's evanescent field on the optical beam side, but it could also be placed to the side of the mechanical beam, as described in Sec.~\ref{sec:discuss}.  This ``TMO'' structure leaves access to the outside of the optical beam unobstructed, eliminating the need for the optical and mechanical design modifications described in Appendix~\ref{app:endfire}.  At the same time, the tuning power of the TMO device is smaller than that of the TOM device due to the greater distance between the tuning beam and the optical slot mode.

To investigate the effect of this alternative tuning beam placement on device performance, we fabricated prototype TMO structures in $\SiN$ without the NEMS actuator electrodes, as shown in Fig.~\ref{fig:tmo}(a). Using a FTW coupled at the side of the optical beam, we characterized the optical and mechanical modes of the device. The optical slot mode at $\approx 962$~nm (Figure~\ref{fig:tmo}(b)) has an intrinsic optical quality factor of $Q_o=120000~\pm~2000$~\cite{mecherrornote}, which is comparable to $Q_o$ previously measured in $\SiN$ slot-mode optomechanical crystals\cite{grutter_slot-mode_2015}, showing that, as with the TOM structure, the presence of the tuning structure does not significantly perturb the optical mode.  The spectrum of the mechanical breathing mode at $\approx 3.03$~GHz is shown in Figure~\ref{fig:tmo}(c).  The breathing mode quality factor $Q_m=2260~\pm~80$,\cite{mecherrornote} which is also comparable to $Q_m$ previously measured in $\SiN$ slot-mode optomechanical crystals\cite{grutter_slot-mode_2015} and is higher than the measured $Q_m$ of the TOM devices of this work. This supports the hypothesis that the asymmetric mechanical beam anchoring for end-fire coupling in the TOM device is the source of its comparatively lower $Q_m$.

\bibliography{references}

\end{document}